# Magnetic Properties and Electronic Structure of Magnetic Topological Insulator MnBi$_2$Se$_4$

Tiancong Zhu,[1*] Alexander J. Bishop,[1*] Tong Zhou,[2] Menglin Zhu,[3] Dante J. O'Hara,[1,4] Alexander A. Baker,[5] Shuyu Cheng,[1] Robert C. Walko,[1] Jacob J. Repicky,[1] Jay A. Gupta,[1] Chris M. Jozwiak,[6] Eli Rotenberg,[6] Jinwoo Hwang,[3] Igor Žutić,[2] and Roland K. Kawakami[1**]

[1]*Department of Physics, The Ohio State University, Columbus, OH 43210, USA*
[2]*Department of Physics, University at Buffalo, Buffalo, NY 14260, USA*
[3]*Department of Materials Science and Engineering, The Ohio State University, Columbus, OH 43210, USA*
[4]*Materials Science and Engineering, University of California, Riverside, CA 92521, USA*
[5]*Lawrence Livermore National Laboratory, Livermore, CA 94550, USA*
[6]*Advanced Light Source, Lawrence Berkeley National Laboratory, Berkeley, CA 94720, USA*

The intrinsic magnetic topological insulators MnBi$_2$X$_4$ (X = Se, Te) are promising candidates in realizing various novel topological states related to symmetry breaking by magnetic order. Although much progress had been made in MnBi$_2$Te$_4$, the study of MnBi$_2$Se$_4$ has been lacking due to the difficulty of material synthesis of the desired trigonal phase. Here, we report the synthesis of multilayer trigonal MnBi$_2$Se$_4$ with alternating-layer molecular beam epitaxy. Atomic-resolution scanning transmission electron microscopy (STEM) and scanning tunneling microscopy (STM) identify a well-ordered multilayer van der Waals (vdW) crystal with septuple-layer base units in agreement with the trigonal structure. Systematic thickness-dependent magnetometry studies illustrate the layered antiferromagnetic ordering as predicted by theory. Angle-resolved photoemission spectroscopy (ARPES) reveals the gapless Dirac-like surface state of MnBi$_2$Se$_4$, which demonstrates that MnBi$_2$Se$_4$ is a topological insulator above the magnetic ordering temperature. These systematic studies show that MnBi$_2$Se$_4$ is a promising candidate for exploring the rich topological phases of layered antiferromagnetic topological insulators.

*equal contributions
**e-mail: kawakami.15@osu.edu

The realization of novel topological quantum states is strongly correlated to the symmetries of different material systems[1-3]. In topological insulators (TIs), the existence of time reversal symmetry (TRS) yields a topologically protected gapless surface state distinguished from its gapped bulk states[4-7]. Breaking the TRS can lead to the quantum anomalous Hall (QAH) effect with dissipationless chiral edge states[8-12], the axion insulator states with quantized magnetoelectric effects[13-15], and Majorana fermions (by coupling to superconductors) obeying non-Abelian statistics[16-18]. Experimentally, the engineering of TRS in TIs has been achieved through doping the TIs with dilute transition metals (Cr, Mn, V etc.)[19], where both QAH[20-22] and axion insulator states[23,24] had been observed at low temperature. However, one drawback of introducing magnetism through dilute doping is that the randomly distributed magnetic dopants can also induce disorder and nonuniformity in the sample, which ultimately limits QAH and axion insulator states from being realized at elevated temperatures[25,26].

Consequently, researchers have increasingly focused on exploring intrinsic magnetic topological insulators, where the magnetic elements are ordered within the crystal lattice[14]. Among such systems, $MnBi_2X_4$ (X = Se or Te) has garnered substantial attention due to predictions of novel topological phases related to layered antiferromagnetic order[27-30]. These materials have a layered van der Waals (vdW) structure with covalently-bonded septuple layers (SL) (X-Bi-X-Mn-X-Bi-X) as the base unit (Fig. 1a), where a magnetic Mn-X layer can be viewed as inserted into each X-Bi-X-Bi-X layer of a topological insulator. Although there have been experimental reports of axion insulator and QAH insulator phases in $MnBi_2Te_4$ [31,32], progress on topological phases in the selenide-based $MnBi_2Se_4$ has been limited by material synthesis. Efforts to synthesize bulk crystals have produced monoclinic $MnBi_2Se_4$, its thermodynamically stable phase[33], as opposed to the desired trigonal structure. In addition, thin film growth has only produced isolated SL within $Bi_2Se_3$ [34] or at its surface[35,36], while the multilayer trigonal $MnBi_2Se_4$ needed to realize new topological phases has remained elusive.

Here, we report the epitaxial growth of multilayer trigonal $MnBi_2Se_4$ and identify ferromagnetism in the single-SL limit, layered antiferromagnetic ordering in multi-SL films, and the presence of topological surface states. We utilize molecular beam epitaxy (MBE) for the growth of $MnBi_2Se_4$ on $Al_2O_3(0001)$ substrates with alternating layer deposition and targeted growth interrupts[37]. Atomic-resolution scanning transmission electron microscopy (STEM) and scanning tunneling microscopy (STM) identify a well-ordered multilayer vdW crystal with SL base units in agreement with the trigonal structure. The magnetic properties of single-layer and multilayer $MnBi_2Se_4$ are measured by superconducting quantum interference device (SQUID) magnetometry. In the single-SL limit, ferromagnetic intralayer coupling with in-plane oriented magnetic moments is observed with SQUID and further confirmed to originate with the Mn atoms (as opposed to impurities) using x-ray magnetic circular dichroism (XMCD).

Layered antiferromagnetic ordering is identified in multilayer MnBi$_2$Se$_4$ by studying the magnetic hysteresis loops and saturation magnetic moment as a function of film thickness. Furthermore, the electronic band structure of MnBi$_2$Se$_4$ is investigated by angle-resolved photoemission spectroscopy (ARPES) and compared with density functional theory (DFT). ARPES measurements of the paramagnetic phase reveals the presence of topological surface states with Dirac dispersion that lie within the bulk bandgap, which demonstrates that MnBi$_2$Se$_4$ is a TI above the magnetic ordering temperature. Our systematic study shows that MnBi$_2$Se$_4$ is a promising candidate for exploring the rich topological phases of layered antiferromagnetic TIs.

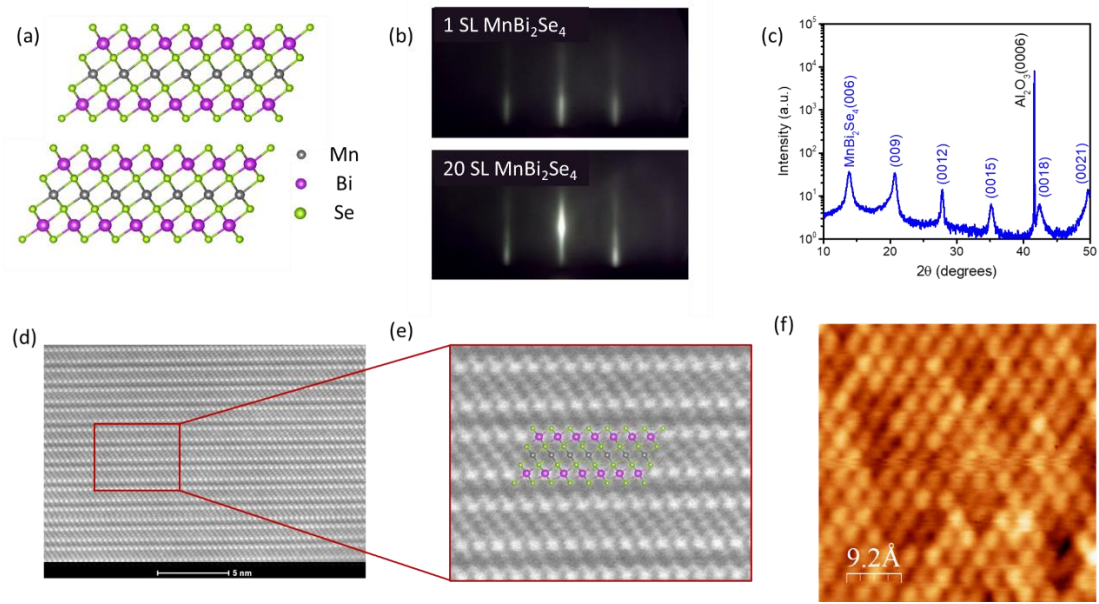

**Figure 1. Structural characterization of multilayer MnBi$_2$Se$_4$ on Al$_2$O$_3$(0001).** (a) The atomic lattice structure of multilayer MnBi$_2$Se$_4$ consists of Se-Bi-Se-Mn-Se-Bi-Se septuple layers (SL) in a van der Waals stack. (b) Streaky RHEED patterns on MnBi$_2$Se$_4$ films indicate a two-dimensional surface structure at thicknesses of 1 SL and 20 SL. (c) θ-2θ x-ray diffraction scans of a 20 SL MnBi$_2$Se$_4$ show that only the substrate(Al$_2$O$_3$(0001)) and the MnBi$_2$Se$_4$ (*00l*) planes are observed. (d) Cross-sectional STEM imaging reveals the van der Waals stacking and atomic-scale structure of the SL (e) Zoomed-in STEM image with atomic lattice overlay that matches with MnBi$_2$Se$_4$. (f) STM imaging reveals the atomic-scale surface structure of a 20 SL MnBi$_2$Se$_4$ film with hexagonal surface structure and an in-plane lattice constant of 4.0 ± 0.1 Å.

Epitaxial MnBi$_2$Se$_4$ films are grown on Al$_2$O$_3$(0001) substrates in a Veeco 930 MBE system (see Methods). With the substrate held at 275 °C and under Se flux, we grow layers of MnBi$_2$Se$_4$ following a three-step process for each SL. First, a quintuple layer of Bi$_2$Se$_3$ is grown by depositing the equivalent of two atomic layers of Bi. Next, we deposit a single atomic layer of Mn and then wait for five minutes under Se flux to allow the Mn to diffuse into the underlying Bi$_2$Se$_3$ template layer to form an individual trigonal MnBi$_2$Se$_4$ SL. The reflection high energy electron diffraction (RHEED) images after each step can be found in Fig. S1 and Fig. S2 of the Supplementary Information (SI). This three-step process is repeated to create multilayer $MnBi_2Se_4$ to the desired thickness, where sharp and streaky RHEED patterns persist throughout the growth (Fig. 1b). The samples are capped with ~10 nm amorphous Se before being removed from the MBE chamber for further characterization.

We perform various structural characterizations of the samples to ensure good crystal quality of the material. θ-2θ x-ray diffraction measurements on a 20 SL sample shows a series of peaks that correspond to the $MnBi_2Se_4$ (*00l*) planes (Fig. 1c). These peaks yield an interplanar spacing of 12.76 Å in the sample, which is distinct from $Bi_2Se_3$ or other Mn-Se compounds. The crystal structure of the MBE-grown $MnBi_2Se_4$ is further investigated with cross-sectional STEM (see Methods). The high-angle annular dark field (HAADF) image in Figure 1d shows the vdW stacking of SLs of the sample, where the atomic structure of each SL in the zoomed-in image (Fig. 1e) agrees very well with the trigonal $MnBi_2Se_4$ structure. The STEM measurement also confirms the continuous growth of multilayer $MnBi_2Se_4$ SLs, where previous studies only show isolated $MnBi_2Se_4$ layers within $Bi_2Se_3$ or at its surface[34,35]. STM measurements on the $MnBi_2Se_4$ surface (Fig. 1f) reveals hexagonal atomic structure of the sample surface with in-plane lattice constant of 4.0 ± 0.1 Å. These structural characterizations confirm the high quality of our MBE-grown $MnBi_2Se_4$ film.

DFT calculations predict that $MnBi_2Se_4$ is magnetic and the magnetic moments are localized on the Mn atoms[27,34,38]. Within the $MnBi_2Se_4$ SL, the magnetic moments are ferromagnetically coupled to each other, while between adjacent $MnBi_2Se_4$ SLs, the magnetic coupling is antiferromagnetic[27]. In order to understand the magnetism in the MBE-grown films, we start by investigating a single SL of $MnBi_2Se_4$. Figure 2a shows the AFM image of a nominally single SL $MnBi_2Se_4$ on $Al_2O_3$(0001). Despite a few 2 SL islands, the single SL $MnBi_2Se_4$ film shows good coverage and uniformity. The height histogram in the inset of Figure 2a shows a predominance of the single layer. We use SQUID magnetometry to study the magnetic properties of the single layer sample at 4 K. Figure 2b compares the magnetization curves of single SL $MnBi_2Se_4$ with magnetic field parallel (in-plane, $H \parallel ab$) or perpendicular (out-of-plane, $H \parallel c$) to the sample surface. When the magnetic field is swept parallel to the sample surface, a clear switch in magnetization is observed at low magnetic field. This shows that single SL $MnBi_2Se_4$ is ferromagnetic, which agrees with the DFT prediction. On the other hand, when the magnetic field is swept perpendicular to the sample surface, a linear magnetization response with applied field is observed instead. The difference in magnetization curves between the in-plane and out-of-plane geometries shows that the magnetic anisotropy of single SL $MnBi_2Se_4$ prefers the magnetization to lie within the surface plane. The magnetic properties are further investigated by varying the sample temperature. As shown in Figure 2c, the ferromagnetism becomes weaker as the temperature is increased above 4 K, and transitions to a non-magnetic state for temperatures above 10 K. A detailed scan of M(T) (Fig. 2c, inset) confirms a Curie temperature, $T_C$, of ~10 K. The origin of the magnetic signal in single SL MnBi$_2$Se$_4$ is investigated by element-specific x-ray magnetic circular dichroism (XMCD) (Fig. 2d). Circularly polarized x-rays are incident upon the sample at 30° from grazing, with magnetic field applied parallel to the x-ray beam. Detection is performed using surface-sensitive total electron yield mode. A difference between left- and right-circularly polarized x-ray absorption is measured at the Mn-L$_{2,3}$ edges at 8 K and 50 kOe, demonstrating that the observed magnetic signal originates with Mn, ruling out the possibility of substrate impurities as the origin of the SQUID results. The combination of SQUID magnetometry and XMCD measurements establish the ferromagnetic ordering in single SL MnBi$_2$Se$_4$ and show that the material is a monolayer 2D magnet. We note here that similar ferromagnetic ordering had been observed in monolayer MnBi$_2$Se$_4$ grown on Bi$_2$Se$_3$ thin films in previous studies[35], but magnetic order was observed up to room temperature. We believe such a discrepancy could be due to the formation of room temperature 2D ferromagnet MnSe$_2$ [39] in their samples, as evidenced by STM images of samples grown under similar conditions[40].

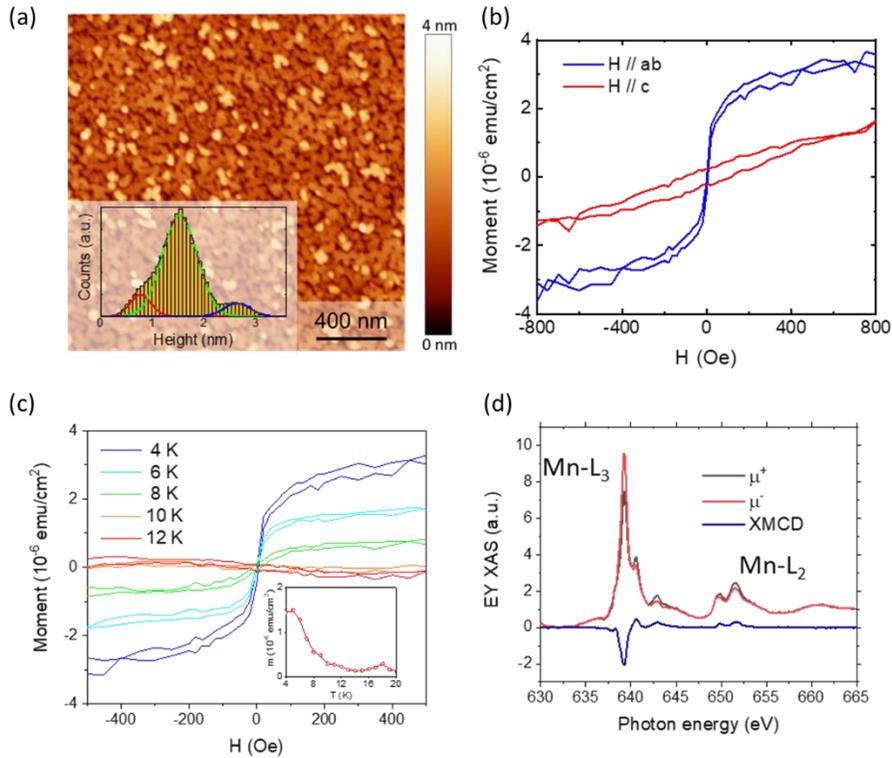

**Figure 2. Ferromagnetism in monolayer MnBi$_2$Se$_4$ on Al$_2$O$_3$(0001).** (a) AFM images of a 1 SL film indicate a relatively flat film with regions of exposed substrate and onset of second layer nucleation. The height histogram (inset) shows the predominance of the monolayer film. (b) SQUID magnetization loops of a 1 SL MnBi$_2$Se$_4$ film at 4 K indicate ferromagnetic ordering with magnetic anisotropy favoring in-plane magnetization. (c) Temperature dependence of in-plane magnetization loops indicate a $T_C$ of ~10 K. The inset shows a detailed temperature dependence with small applied field of 50 Oe. (d) Element-specific XMCD measurement at 5 T and 8 K, where x-

ray absorption spectra at the Mn L-edge for μ+ circular polarization (black curve) and μ- circular polarization (red curve) exhibits an asymmetry (blue curve) that confirms a net magnetic moment from Mn.

After confirming the ferromagnetic ordering of 1 SL of $MnBi_2Se_4$, we study the magnetism in few-layer films to understand the interlayer magnetic coupling. Figure 3 shows the SQUID magnetometry measurement on a 7 SL $MnBi_2Se_4$ film. In the low field range, the magnetic behavior of the few-layer $MnBi_2Se_4$ film is similar to that of the 1 SL sample. Sweeping in-plane magnetic field produces a clear hysteresis loop, while an out-of-plane magnetic field sweep yields an almost linear response in magnetization (Fig. 3a). This further confirms magnetic anisotropy favoring magnetic moments to lie within the film plane. M(H) scans for different temperatures of the 7 SL film (Fig. 3b) also show a magnetic ordering temperature of ~10 K, similar to the 1 SL sample.

However, while the out-of-plane hysteresis loop appears to close when the external magnetic field is above ~500 Oe (Fig. 3a, blue curve), the magnetization of the sample has not reached saturation. This behavior is different from the single SL sample. In order to understand this, we measure the magnetic moment (normalized by the sample area) of the 7 SL sample up to high magnetic field (Fig. 3c). With increasing field, the net magnetic moment keeps increasing with higher external magnetic field, and eventually saturates at ~25 kOe. The saturation magnetic moment with H > 25 kOe is about six times higher compared to the total moment of the sample when the hysteresis loop closes (~500 Oe).

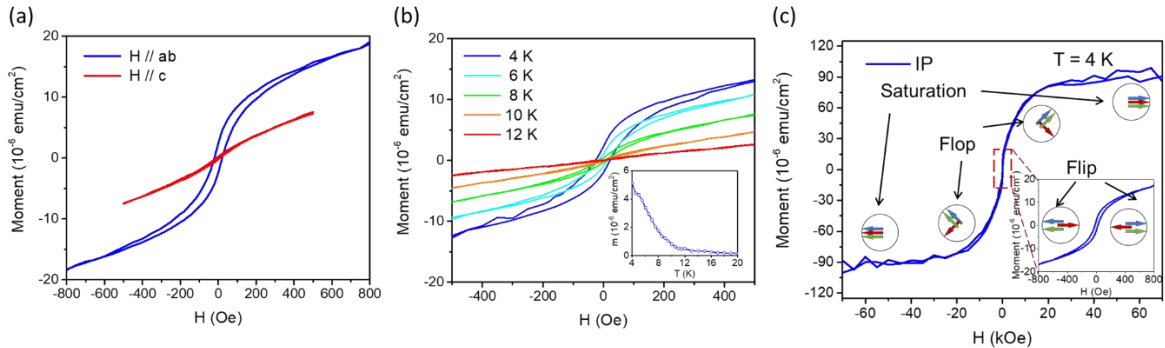

**Figure 3. Low-field magnetic properties of 7 SL $MnBi_2Se_4$ on $Al_2O_3$(0001).** (a) A comparison of in-plane vs. out-of-plane SQUID magnetization loops at 4 K suggests an easy-plane magnetic anisotropy. (b) In-plane SQUID magnetization loops at different temperatures exhibit hysteresis at 4 K which disappears at 10 K. The inset shows a detailed temperature dependence of magnetization at a fixed field of 50 Oe. (c) In-plane SQUID magnetization loops of 7 SL $MnBi_2Se_4$ at 4 K exhibit hysteresis at low field due to global spin flip with uncompensated magnetic moments (inset), a gradual increase of magnetization due to spin flop, and saturation as moments align at high field. The blue/red/green arrows represent the magnetization direction of three different layers (instead of seven layers for clarity) of $MnBi_2Se_4$ viewed along the (001) direction.

The hysteretic behavior at low magnetic field and the saturation of net magnetic moment at high magnetic field is consistent with the predicted layered antiferromagnetic ordering in $MnBi_2Se_4$, as illustrated in Figure 3c. In a layered antiferromagnet at zero field, the magnetization of adjacent layers is antiparallel to each other due to interlayer antiferromagnetic coupling. This results in a zero net magnetic moment when the number of layers is even, and leaving an uncompensated magnetic layer and non-zero net magnetic moment when the number of layers is odd. Under a small magnetic field, the magnetic moments in the 7 SL sample can undergo a 'flip' process, where all the magnetic moments change to the opposite direction. This process maintains the antiferromagnetic coupling between each layer, while aligning the uncompensated net magnetic moment with the external field to minimize the total energy. However, at sufficiently high magnetic field (>25 kOe) all the magnetic moments align parallel (i.e. saturate) when the Zeeman energy overcomes the interlayer antiferromagnetic coupling. In the intermediate field regime, the gradual increase of the magnetization toward saturation, as opposed to an abrupt switching at a particular field (i.e split hysteresis loop)[41,42], suggests a spin-flop antiferromagnetic alignment with canted moments lying in the sample plane. Thus, the change of net magnetic moment in the 'flip' process should not exceed the magnetic moment in a single SL because the antiparallel moments cancel and only the uncompensated moments contribute[43-45]. On the other hand, the saturation magnetic moment should scale linearly with the sample thickness because all moments are parallel.

To confirm this hypothesis, we perform magnetometry measurements on multiple $MnBi_2Se_4$ films of different thicknesses. To assess the scaling with thickness, we plot the total magnetic moment (normalized by sample area) for representative samples (Fig. 4a) and find that the total saturation moment at high field ($m_{High}$) increases with sample thickness. This behavior is expected because the magnetic moments are aligned in the saturated state and their quantity increases linearly with film thickness. In order to investigate the thickness dependence of the low field 'flip' process, we quantify the net moment (normalized by sample area) associated with the low field flip ($m_{Low}$) by linear-extrapolation of the closed part of the low field hysteresis loop back to zero magnetic field (inset of Fig. 4a). Figure 4b summarizes the thickness dependence of $m_{High}$ and $m_{Low}$ across 10 different samples. The expected linear scaling of $m_{High}$ with thickness is observed and the slope of 12.7 ± 1.9 emu/cm$^2$ per SL corresponds to 2.0 ± 0.3 $\mu_B$ per Mn. On the other hand, $m_{low}$ shows relatively little variation across the whole thickness range. The measured $m_{Low}$ of most of the samples are within the magnetization per single SL $MnBi_2Se_4$ extracted from the slope of the fitted line in Figure 4b. Ideally, the $m_{Low}$ value should oscillate on-and-off with the sample thickness changing from odd to even. Such behavior is not clearly observed in our experimental data, possibly due to atomic-scale thickness fluctuations producing mixed even-odd thicknesses in our epitaxial films. Nevertheless, the observed behaviors of both $m_{Low}$ and $m_{High}$ agree well with the expectations for layered antiferromagnetism in $MnBi_2Se_4$.

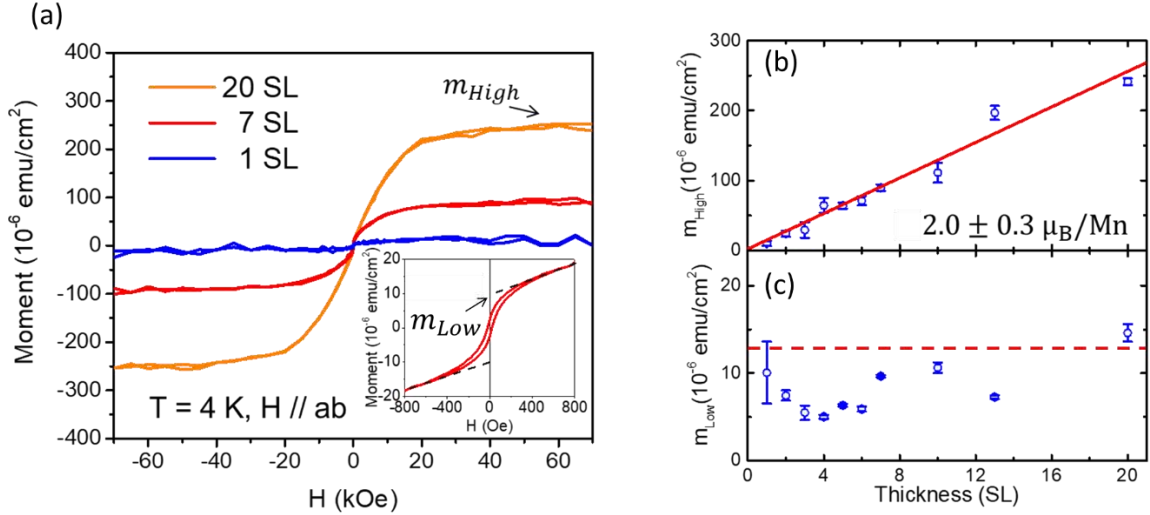

**Figure 4. Evidence for layered antiferromagnetism in MnBi$_2$Se$_4$ on Al$_2$O$_3$(0001).** (a) Comparison of in-plane magnetization loops of 1 SL, 7 SL, and 20 SL films show an increase of the saturation moment per area ($m_{High}$) with increasing thickness. All loops exhibit low-field hysteresis, where we define the extrapolated remanence ($m_{Low}$) as shown in the inset. (b) A linear dependence of $m_{High}$ with film thickness is consistent with aligned magnetic moments in high field. (c) $m_{Low}$ showing little variation with film thickness is consistent with a low field antiferromagnetic state with net uncompensated moments or surface ferromagnetism. The red dashed line marks the magnetic moment for 1 SL MnBi$_2$Se$_4$ based on 2.0 μ$_B$/Mn as determined by the slope in (b). Note: $m_{Low}$ for the 1 SL sample in (c) is determined using the saturation value at high magnetic field (same method as $m_{High}$), due to no obvious hysteresis loop observed at low field (see Fig. 2(b))

Finally, in order to investigate the electronic and topological properties, we perform ARPES measurements (see Methods) on a 7 SL MBE grown MnBi$_2$Se$_4$ film and compare with DFT calculations (see Methods). The ARPES measurements are performed at 80 K, where the sample is paramagnetically ordered. Simulating the paramagnetic order using DFT calculations is challenging. However, a nonmagnetic case can be a good approximation to capture the key electronic structure at the Γ point of the paramagnetic one, which has been verified in a similar system MnBi$_2$Te$_4$ [46]. Thus, we calculate the bands of a nonmagnetic MnBi$_2$Se$_4$ crystal as shown in Figure 5a. The key message here is the topological property of the bands at Γ point. The calculated topological invariant Z$_2$ of 1 shows a nontrivial gap of ~0.2 eV at the Γ point, indicating that gapless Dirac surface states exist in the MnBi$_2$Se$_4$ surface, which is confirmed by the surface states calculation. Figure 5b shows the energy-momentum relation along the M-Γ-M direction measured by ARPES (other cuts are shown in the SI, Fig. S3), which agrees well with the DFT calculation. To make a reliable identification of the bulk bands, we overlay the DFT calculation with the experimental data (Fig. 5c), where we have taken the second derivative of the ARPES data along the energy axis to improve the contrast of the energy bands. The excellent overlap of the DFT calculation with the experimental spectra provides a confident assignment of the bulk conduction and valence bands.

Based on this assignment of bands, we identify the positions of the conduction band minima (CBM) and valence band maxima (VBM) in the ARPES spectra.

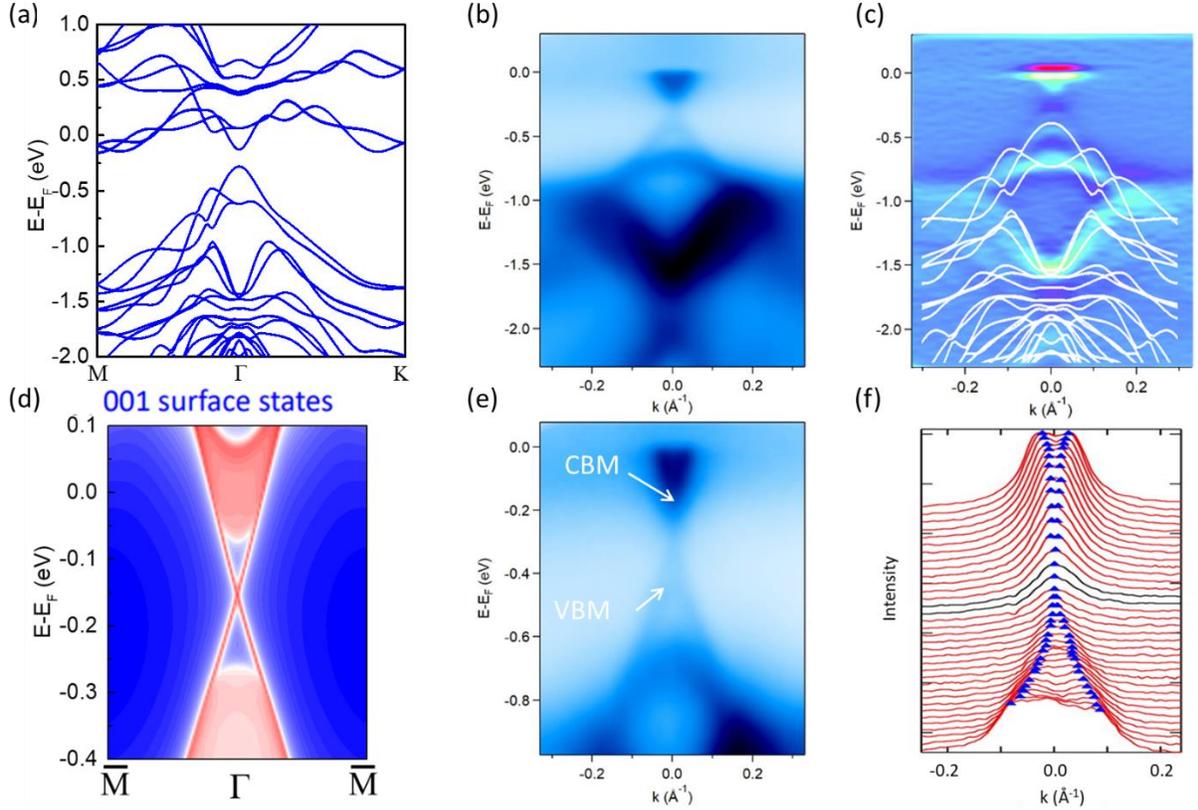

**Figure 5. Bulk and surface electronic states.** (a) DFT calculation of the bulk states with Mn treated as nonmagnetic. (b) ARPES measurement of a 7 SL MnBi$_2$Se$_4$ film along M-Γ-M with 20 eV photon energy and sample temperature of 80 K. (c) Overlay of the DFT calculation and 2$^{nd}$-derivative of the ARPES data in (b) shows a good agreement of calculation with the experiment. (d) DFT-calculated (001) surface states of nonmagnetic MnBi$_2$Se$_4$. (e) Zoomed-in view of the ARPES data that highlights the Dirac surface state. CBM and VBM indicate the bulk conduction band minimum and valence band maximum. (f) A series of momentum distribution curves showing the evolution of the Dirac cone for energies ranging from -0.75 to 0.0 eV. Local maxima are indicated by blue triangles. Black curves indicate the Dirac point.

The existence of a Dirac-like surface state in nonmagnetic MnBi$_2$Se$_4$ is clearly observed in zoomed-in ARPES spectra (Fig. 5e), where the linearly dispersing bands are connecting the CBM and VBM of MnBi$_2$Se$_4$. To investigate the nature of these surface states, we perform DFT calculations on a semi-infinite slab. The calculation confirms the presence of a topological surface state with Dirac dispersion (red lines in Fig. 5d) in addition to the bulk states, which appear as the solid red regions of Figure 5d when projected onto the (001) plane. To investigate the surface state in the experimental data, we plot the momentum distribution curves for a series of energies that traverse the bulk band gap (Fig. 5f).

We observe a crossing of the local maxima (indicated by the blue triangles) which signifies the Dirac point and confirms the gapless nature of the surface state within our energy resolution.

In conclusion, we have developed the MBE synthesis of multilayer trigonal $MnBi_2Se_4$. STEM and STM imaging confirm the high quality of the material and identify the atomic-scale structure of the septuple layer – the building block of the van der Waals crystal. Magnetic measurements indicate that the single layer $MnBi_2Se_4$ is a monolayer 2D ferromagnet with a Curie temperature of ~10 K, while multilayer $MnBi_2Se_4$ is a layered antiferromagnet. This layer-dependent magnetic character of $MnBi_2Se_4$ is also consistent with our DFT studies. ARPES measurements and DFT calculations confirm the presence of a topological Dirac surface state, which shows that $MnBi_2Se_4$ in the paramagnetic state is a topological insulator.

With the realization of this high-quality magnetic TI we expect several interesting directions for future research in van der Waals heterostructures. In the normal state, magnetic TI/TI heterostructures could enable peculiar tunneling planar Hall effect, which attains its maximum signal for the in-plane magnetization along the current direction, when the usual planar and other Hall effect vanish[47,48]. In the superconducting heterostructures, such magnetic TIs could provide both magnetic proximity effects and synthetic spin-orbit coupling through magnetic textures to support spin-triplet superconductivity, an important element for superconducting spintronics[49,50] and tunable Majorana bound states[48,51,52].

**Methods**

**Molecular beam epitaxy.** Epitaxial growth is performed in a Veeco 930 MBE chamber. $Al_2O_3(0001)$ substrates (MTI Corporation) are annealed in air at 1000 °C for 3 hours then annealed under ultrahigh vacuum (UHV) at 600 °C for 30 minutes. Films are grown using elemental Mn (99.98%, Alfa Aesar) and Bi (99.998%, Alfa Aesar) deposited from effusion cells and Se (99.999%, Alfa Aesar) deposited from a valved cracker (900 °C in cracking zone, 290 °C in the bulk zone). Typical growth rates are 2.5 Å/min for Mn and 1.7 Å/min for Bi, while the Se flux is maintained at overpressure (~1 x $10^{-6}$ torr) with flux ratios of ~100 for Se:Mn and Se:Bi as measured by the beam flux monitor (ion gauge). Similar to Levy *et al.*[53], we initiate the growth with a submonolayer amorphous $Bi_2Se_3$ layer at 140 °C, then re-evaporate the layer at 360 °C prior to the $MnBi_2Se_4$ film growth. This procedure helps to reduce the formation of twinning-domains and increases the surface smoothness of the $MnBi_2Se_4$ film. After growth, the sample is capped with 10 nm amorphous Se to preserve the sample as it is transferred to measurement systems.

**Scanning transmission electron microscopy.** Cross-section TEM specimens of $MnBi_2Se_4$ were made using focused ion beam (Thermofisher Helios), which were subsequently cleaned using a low energy ion mill (Fischione Nanomill) with the beam energy of 500 eV. STEM was performed using a Thermofisher Titan STEM operated at 300 kV. Atomic resolution high angle annular dark field (HAADF) images were acquired at the scattering range between 80 and 300 mrad, where the scattering intensity mostly depends on the Z-number of the atoms.

**X-ray magnetic circular dichroism.** The 1 SL film was capped with a thin (~3 nm) layer of amorphous Se to prevent oxidation and transported to beamline 4-ID-C of the Advanced Photon Source (APS). The sample was mounted and its surface contacted using silver paint. XMCD measurements were performed at the LHe cryostat's base temperature of 8 K, as measured by a thermocouple close to the sample. A magnetic field of 50 kOe was applied parallel to the beam, and x-rays were incident upon the sample at 30° from grazing incidence (60° from normal). X-ray absorption spectra (XAS) were gathered in surface-sensitive total electron yield detection, using positive and negative polarizations, and with positive and negative fields. Spectra were normalized to the incident beam intensity, measured with a gold mesh just upstream of the magnet chamber, and energy calibration was performed using a portion of the beam split off onto a MnO standard. A linear background was subtracted from the XAS spectra, and the pre-edge to post-edge step scaled to 1, to account for transitions into the continuum. XMCD was calculated as the difference between matched pairs (fixed polarization, switched field and switched polarization, fixed field) to ensure proper correction of incident beam effects.

**Angle-resolved photoemission spectroscopy.** The 7 SL $MnBi_2Se_4$ film is transported in an air-tight container under inert gas ($N_2$) to the Microscopic and Electronic Structure Observatory (MAESTRO) at the Advanced Light Source (ALS). The sample is the loaded into UHV and given a mild anneal (~420 K) to remove the amorphous Se capping layer and then transferred into the microARPES endstation where it is cooled to ~80 K. The focused synchrotron beam has a spot size on the order of 10 μm for photon energy of 20 eV used to obtain the data seen in Figure 5. Photoemission spectra are collected using a Scienta R4000 electron analyzer equipped with custom-built deflectors that allow spectra over the first Brillouin zone to be taken on a single spot without moving the sample.

**Density functional theory.** The electronic band structures of $MnBi_2Se_4$ crystal were computed using the general potential linearized augmented plane-wave (LAPW) method[54] as implemented in the WIEN2K code[55]. Spin-orbit coupling is included as a second vibrational step using scalar-relativistic eigenfunctions as the basis after the initial calculation is converged to self-consistency. The lattice structure of $MnBi_2Se_4$ adopted in the calculations is shown in Figure S4a (SI). The optimized lattice constants of $a = b = 3.93$ Å and $c = 36.78$ Å are in a good agreement with our STM measurements. Figure S4b (SI) shows the high symmetry points of the Brillouin zone (BZ). The convergence of the calculations regarding the size of the basis set is achieved using an $R_{MT} * K_{max}$ value of 7, where RMT is the smallest atomic sphere radius in the unit cell and $K_{max}$ is the magnitude of the largest K wave vector inside the first BZ. The local density approximation (LDA) is used to describe the exchange and correlation functional. To take into account the correlation effects of Mn $3d$ electrons, the Hubbard U parameter is taken as 4 eV in the LDA+U calculations[56]. The Monkhorst-Pack k-grid of 12 × 12 × 3 is adopted for the first BZ integral and the convergence criterion for the charge difference was less than $0.001e$ per unit cell. The surface states are calculated using the iterative Green's function technique as implemented in WannierTools package[57] based on the tight-binding model constructed by the maximally localized Wannier functions[58].

**Acknowledgements**. We thank J. Freeland and J-B. Forien for support on the XMCD measurements, M. R. Brenner for support on the MBE growth, and A. Bostwick and R. A. Bennett for assistance with the ARPES measurements. This work was primarily supported by the Department of Energy (DOE) Basic Energy Sciences under Grant No. DE-SC0016379. AJB and RKK acknowledge partial support from AFOSR MURI 2D MAGIC (FA9550-19-1-0390). Part of this work was performed under the auspices of the U.S. DOE by Lawrence Livermore National Laboratory (LLNL) under Contract No. DE-AC52-07NA27344, and was supported by the LLNL-LDRD program under Project No. 19-LW-028. Portions of this work were performed at Sector 4 of the APS, Argonne National Laboratory. The APS is a U.S. DOE Office of Science User Facility operated for the DOE Office of Science by ANL under Contract No. DE-AC02-06CH11357. This research used resources of the Advanced Light Source, a DOE Office of Science User Facility under contract no. DE-AC02-05CH11231. MZ and JH acknowledge partial support from the Center for Emergent Materials, an NSF MRSEC, under Grant No. DMR-1420451. TZhou and IZ were supported by U.S. DOE Office of Science, Basic Energy Sciences under Grant No. DE- SC0004890 and the UB Center for Computational Research.

**Author contributions.** RKK and TZhu conceived the study. TZhu and AJB synthesized the material with the help of DJO. TZhu performed the magnetometry measurement and data analysis with the help of AJB and DJO. MZ performed the STEM measurement. TZhu and RCW performed the STM measurement with the help of JJR. DJO and AAB performed the XMCD measurement. TZhu, AJB, SC, and RKK performed the ARPES measurements with CMJ and ER. TZhou and IZ performed DFT calculations and theoretical analysis. All authors contributed to the interpretation of the results and preparation of the manuscript.

36. Okuyama, Y., Ishikawa, R., Kuroda, S. & Hirahara, T. Role of hybridization and magnetic effects in massive Dirac cones: Magnetic topological heterostructures with controlled film thickness. *Appl. Phys. Lett.* **114**, 051602 (2019).
37. Gong, Y., Guo, J., Li, J., Zhu, K., Liao, M., Liu, X., Zhang, Q., Gu, L., Tang, L., Feng, X., Zhang, D., Li, W., Song, C., Wang, L., Yu, P., Chen, X., Wang, Y., Yao, H., Duan, W., Xu, Y., Zhang, S.-C., Ma, X., Xue, Q.-K. & He, K. Experimental Realization of an Intrinsic Magnetic Topological Insulator. *Chin. Phys. Lett.* **36**, 076801 (2019).
38. Hou, Y. & Wu, R. Axion Insulator State in a Ferromagnet/Topological Insulator/Antiferromagnet Heterostructure. *Nano Lett.* **19**, 2472-2477 (2019).
39. O'Hara, D. J., Zhu, T., Trout, A. H., Ahmed, A. S., Luo, Y. K., Lee, C. H., Brenner, M. R., Rajan, S., Gupta, J. A., McComb, D. W. & Kawakami, R. K. Room Temperature Intrinsic Ferromagnetism in Epitaxial Manganese Selenide Films in the Monolayer Limit. *Nano Lett.* **18**, 3125-3131 (2018).
40. Noesges, B. A., Zhu, T., Repicky, J. J., Yu, S., Yang, F., Gupta, J. A., Kawakami, R. K. & Brillson, L. J. Chemical Migration and Dipole Formation at van der Waals Interfaces between Magnetic Transition Metal Chalcogenides and Topological Insulators. *in preparation*.
41. Qiu, Z. Q., Pearson, J., Berger, A. & Bader, S. D. Short-period oscillations in the interlayer magnetic coupling of wedged Fe(100)/Mo(100)/Fe(100) grown on Mo(100) by molecular-beam epitaxy. *Phys. Rev. Lett.* **68**, 1398-1401 (1992).
42. Huang, B., Clark, G., Navarro-Moratalla, E., Klein, D. R., Cheng, R., Seyler, K. L., Zhong, D., Schmidgall, E., McGuire, M. A., Cobden, D. H., Yao, W., Xiao, D., Jarillo-Herrero, P. & Xu, X. Layer-dependent ferromagnetism in a van der Waals crystal down to the monolayer limit. *Nature* **546**, 270-273 (2017).
43. Cui, J., Shi, M., Wang, H., Yu, F., Wu, T., Luo, X., Ying, J. & Chen, X. Transport properties of thin flakes of the antiferromagnetic topological insulator $MnBi_2Te_4$. *Phys. Rev. B* **99**, 155125 (2019).
44. Yan, J. Q., Zhang, Q., Heitmann, T., Huang, Z., Chen, K. Y., Cheng, J. G., Wu, W., Vaknin, D., Sales, B. C. & McQueeney, R. J. Crystal growth and magnetic structure of $MnBi_2Te_4$. *Phys. Rev. Mater.* **3**, 064202 (2019).
45. Lee, S. H., Zhu, Y., Wang, Y., Miao, L., Pillsbury, T., Yi, H., Kempinger, S., Hu, J., Heikes, C. A., Quarterman, P., Ratcliff, W., Borchers, J. A., Zhang, H., Ke, X., Graf, D., Alem, N., Chang, C.-Z., Samarth, N. & Mao, Z. Spin scattering and noncollinear spin structure-induced intrinsic anomalous Hall effect in antiferromagnetic topological insulator $MnBi_2Te_4$. *Phys. Rev. Res.* **1**, 012011 (2019).
46. Hao, Y.-J., Liu, P., Feng, Y., Ma, X.-M., Schwier, E. F., Arita, M., Kumar, S., Hu, C., Lu, R. e., Zeng, M., Wang, Y., Hao, Z., Sun, H.-Y., Zhang, K., Mei, J., Ni, N., Wu, L., Shimada, K., Chen, C., Liu, Q. & Liu, C. Gapless Surface Dirac Cone in Antiferromagnetic Topological Insulator $MnBi_2Te_4$. *Phys. Rev. X* **9**, 041038 (2019).
47. Tang, H. X., Kawakami, R. K., Awschalom, D. D. & Roukes, M. L. Giant Planar Hall Effect in Epitaxial (Ga,Mn)As Devices. *Phys. Rev. Lett.* **90**, 107201 (2003).
48. Fatin, G. L., Matos-Abiague, A., Scharf, B. & Žutić, I. Wireless Majorana Bound States: From Magnetic Tunability to Braiding. *Phys. Rev. Lett.* **117**, 077002 (2016).
49. Linder, J. & Robinson, J. W. A. Superconducting spintronics. *Nat. Phys.* **11**, 307-315 (2015).
50. Martínez, I., Högl, P., González-Ruano, C., Cascales, J. P., Tiusan, C., Lu, Y., Hehn, M., Matos-Abiague, A., Fabian, J., Žutić, I. & Aliev, F. G. Interfacial Spin-Orbit Coupling: A Platform for Superconducting Spintronics. *Phys. Rev. Appl.* **13**, 014030 (2020).
51. Zhou, T., Mohanta, N., Han, J. E., Matos-Abiague, A. & Žutić, I. Tunable magnetic textures in spin valves: From spintronics to Majorana bound states. *Phys. Rev. B* **99**, 134505 (2019).

# Supplementary Information

# Magnetic Properties and Electronic Structure of Magnetic Topological Insulator $MnBi_2Se_4$

Tiancong Zhu,[1] Alexander J. Bishop,[1] Tong Zhou,[2] Menglin Zhu,[3] Dante J. O'Hara,[1,4] Alexander A. Baker,[5] Shuyu Cheng,[1] Robert C. Walko,[1] Jacob J. Repicky,[1] Jay A. Gupta,[1] Chris M. Jozwiak,[6] Eli Rotenberg,[6] Jinwoo Hwang,[3] Igor Žutić,[2] and Roland K. Kawakami[1]

[1]*Department of Physics, The Ohio State University, Columbus, OH 43210, USA*
[2]*Department of Physics, University at Buffalo, Buffalo, NY 14260, USA*
[3]*Department of Materials Science and Engineering, The Ohio State University, Columbus, OH 43210, USA*
[4]*Materials Science and Engineering, University of California, Riverside, CA 92521, USA*
[5]*Lawrence Livermore National Laboratory, Livermore, CA 94550, USA*
[6]*Advanced Light Source, Lawrence Berkeley National Laboratory, Berkeley, CA 94720, USA*

**Contents:**

1. **Evolution of RHEED images during $MnBi_2Se_4$ growth**
2. **Additional ARPES data**
3. **Unit cell and Brillouin zone of $MnBi_2Se_4$**

## 4. Evolution of RHEED images during MnBi$_2$Se$_4$ growth

The RHEED patterns seen in Figure 1b of the main paper were taken *in situ* with a standard RHEED system from STAIB Instruments. Figure S1 shows RHEED pattern images captured after each growth stage for the first three septuple layers (SLs). Each panel is labeled with a number and letter to indicate the growth layer and growth step respectively. As shown in the figure, letter "A" denotes an image taken immediately after the deposition of two atomic layers of Bi with the Se shutter open, called step A. Letter "B" labels images taken immediately after the deposition of one atomic layer of Mn also under Se flux, known as step B. Finally, letter "C" indicates images taken after the designed growth interrupt with Se flux, called step C.

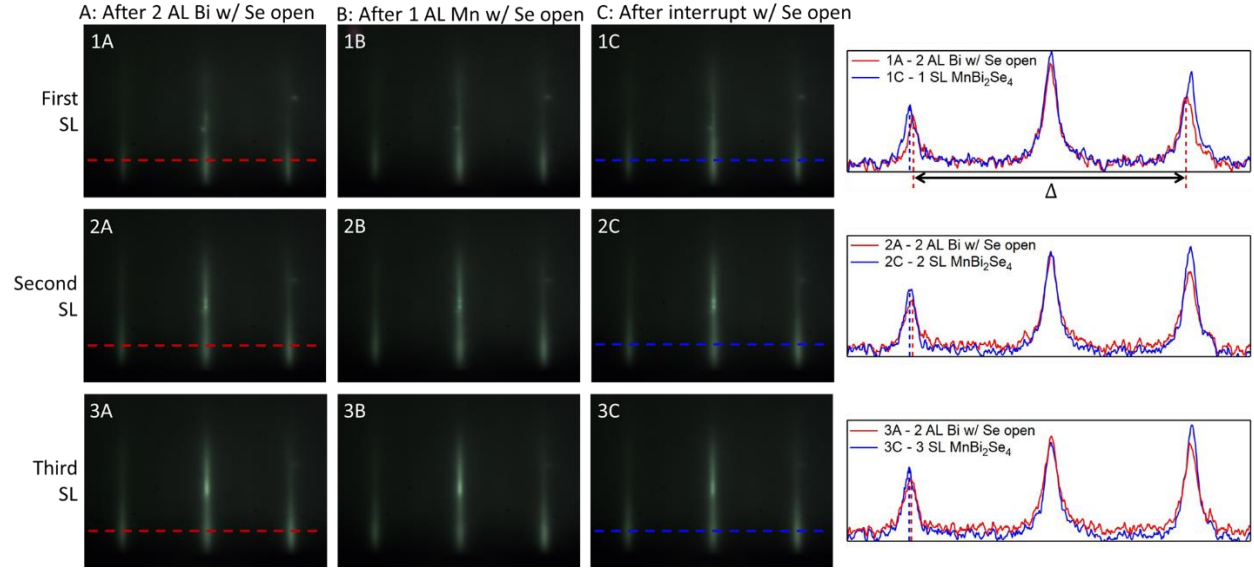

**Figure S1. RHEED evolution of MnBi$_2$Se$_4$ growth.** RHEED patterns taken along the [11$\bar{2}$0] direction after each step of the growth process for the first three septuple layers of MnBi$_2$Se$_4$ growth. Each panel is given a label with the number indicating the septuple layer in progress and the letter indicating which step the growth process is at as described in the text. The graphs to the right of the image display intensity profiles for the first and last growth step of each septuple layer.

Intensity line cuts of RHEED images from step A and step C of each layer and can be seen on the right side of Figure S1. These profiles show that the spacing of the RHEED streaks, denoted by Δ, changes after the first and last steps of the growth. It can be seen that the Δ measured after the designed growth interrupt (step C) is larger than the Δ measured immediately after the deposition of Bi (step A) at each septuple layer. This Δ was measured after steps A and C for the first five SLs of a growth and analyzed in Figure S2. Each point in Figure S2 is a calculated percent difference between the initial 2 AL of Bi deposited for the first SL (labeled 1A in Figure S1) and the growth step being analyzed. Red dots indicate RHEED spacing measured after step A of a given SL growth while blue dots represent the spacing measured after the completion of a full SL. Note that for each SL, the Δ measured after step A is smaller than the Δ measured after step C. This indicates the formation of Bi$_2$Se$_3$ after each A step, and then a transition into MnBi$_2$Se$_4$ upon completion of all growth stages. *In situ* analysis of this spacing is an early indication of successful growth of MnBi$_2$Se$_4$ SLs. Also note that the initial 2 AL of Bi (step 1A) compared to the spacing measured after the completion of the first SL (step 1C) gives a percent difference of ~2.5%. This agrees well with the percent difference between the accepted value of the $Bi_2Se_3$ in-plane lattice constant of 4.14 Å and the measured $MnBi_2Se_4$ lattice constant of 4.0 Å.

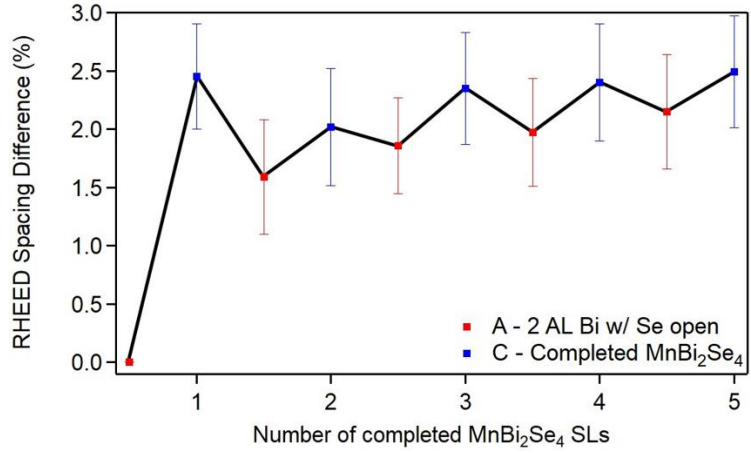

**Figure S2. Evolution of RHEED spacing during growth.** Percent difference of RHEED spacing at first and last steps of growth during the first five septuple layers.

2. Additional ARPES data

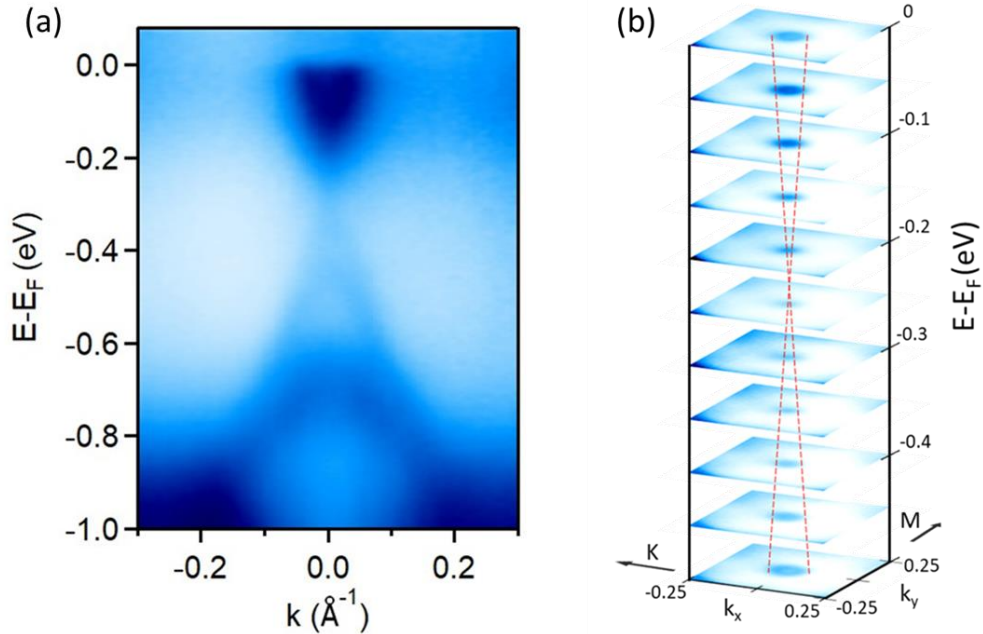

**Figure S3. Additional ARPES data.** (a) ARPES measurement of a 7 SL $MnBi_2Se_4$ film along the K-Γ-K direction. (b) Constant energy contours of the 7 SL film at different energies. All measurements were taken with a photon energy of 20 eV and sample temperature of 80 K.

## 3. Unit cell and Brillouin zone of MnBi$_2$Se$_4$

Figure S4a shows the atomic positions within the unit cell (black rectangle) used for the DFT calculations of MnBi$_2$Se$_4$. Figure S4b indicates the locations of the high symmetry points in the bulk and (001) surface Brilliouin zones.

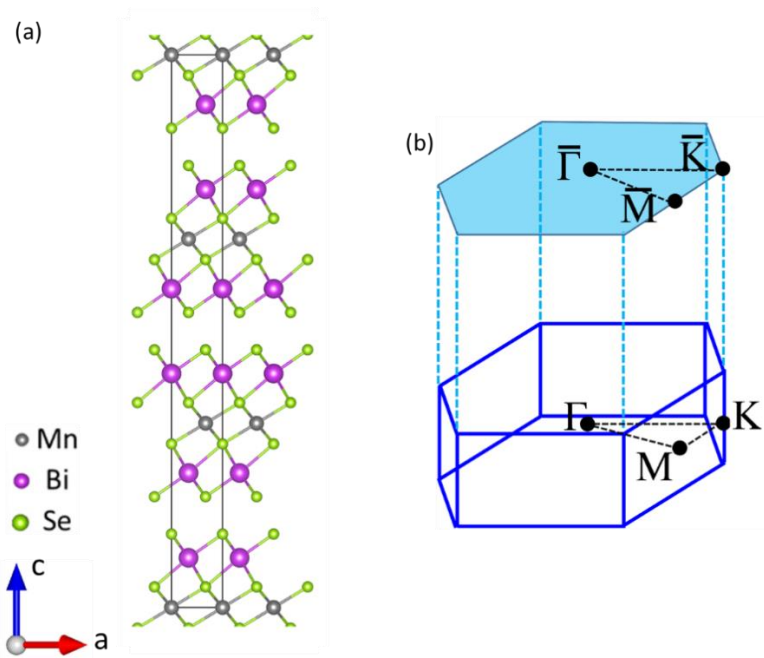

**Figure S4.** (a) The lattice structure of MnBi$_2$Se$_4$ crystal. The black lines indicate the unit cell adopted in the DFT calculations. (b) The bulk Brillouin zone (blue wireframe) and (001) surface Brillouin zone (shaded light blue). The high symmetry points used in Figure 5 of the main text are indicated here.